\documentclass[twocolumn]{aa}
\usepackage{graphicx}
\begin{document}
\title{Systematic motions in the galactic plane found in the Hipparcos Catalogue
 using Herschel's Method}
\author{Carlos Abad       \inst{1,2} \&
        Katherine Vieira  \inst{1,3}}
 
\offprints{C. Abad}
 
\institute{(1) Centro de Investigaciones de Astronom\'{\i}a
               CIDA, 5101-A M\'erida, Venezuela\\ \email{abad@cida.ve}\\
           (2) Grupo de Mec\'anica Espacial, Depto. de
               F\'{\i}sica Te\'orica, Universidad de Zaragoza.
               50006 Zaragoza, Espa\~na\\
	   (3) Department of Astronomy, Yale University, P.O. Box 208101
               New Haven, CT 06520-8101, USA\\ \email{vieira@astro.yale.edu}}

\date{\today}

\authorrunning{C. Abad \& K. Vieira: }
\titlerunning{Systematic motions in the galactic plane}

\abstract{
Two motions in the galactic plane have been detected
and characterized, based on the determination of a common systematic
component in Hipparcos catalogue proper motions. The procedure is based only on
positions, proper motions and parallaxes, plus a 
special algorithm which is able to reveal systematic
trends.  Our results come from two stellar samples.
Sample 1 has 4566 stars and defines a motion
of apex $(l,b)=(177^o8,3^o7)\pm(1^o5,1^o0)$ and space velocity
$V=27\pm 1$ km/s. Sample 2 has 4083 stars and defines a motion
of apex $(l,b)=(5^o4,-0^o6)\pm (1^o9,1^o1)$ and space velocity
$V=32\pm 2$ km/s. Both groups are distributed all over the sky and
cover a large variety of spectral types, which means that they do not
belong to a specific stellar population. Herschel's method is used 
to define the initial samples of stars and later 
to compute the common space velocity. The intermediate process
is based on the use of a special algorithm to determine
systematic components in the proper motions.
As an important contribution, this paper sets out a new way to study
the kinematics of the solar neighborhood, in the search for streams,
associations, clusters and any other space motion shared by a
large number of stars, without being restricted by the availability
of radial velocities.

  }

\maketitle

\section{Introduction}

\begin{figure*}
\begin{center}
\includegraphics[width=15cm,height=8cm,angle=0]{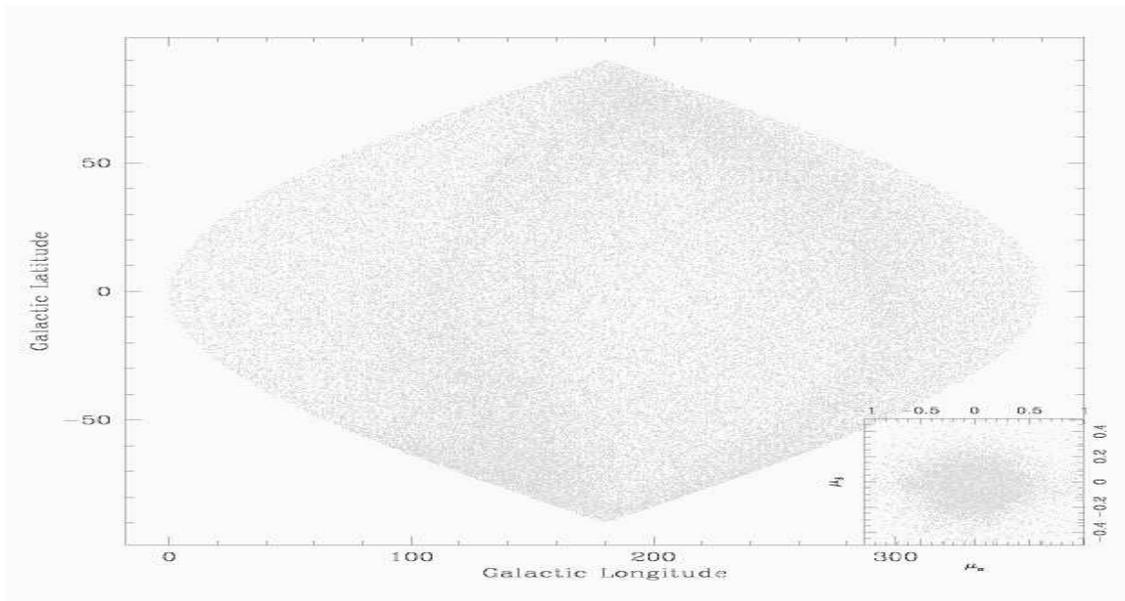}
\caption{A comparative view of 
the Convergent Point Method (CPM) and the Vector-Point Diagram, applied
to Hipparcos catalogue. CPM is plotted using the poles of the great circles.
An area of enhanced density of poles stands out, on which is possible to fit a great
circle.}
\end{center}
\end{figure*}

The proper motion is a linear representation of the real motion of
the stars as it is observed by us. It can be defined
as the projection of the instantaneous velocity with modulus and
direction constant over the celestial sphere. It is inversely
proportional to the distance from the star and its certainty depends
directly on the accuracy of the individual stellar positions
con\-tri\-bu\-ting to it. Obviously, most measured
proper motions are small because the distance-to-speed ratio is
large.

Hipparcos is an astrometric catalogue with accurate
positions and excellent proper motions, measured over a short interval
of time. It includes positions and proper motions,
important information about the stars such as parallax, color and magnitude,
but it lacks radial velocities, which are needed to compute space
velocities. This situation is regretted by some authors as an important
obstacle to study the 3D ste\-llar motions in the solar neighborhood.
In this paper we emphasize that when there is a common component of
motion in a sample of stars, and if the proper motions are properly
analyzed, more information can be obtained about the 3D space motion,
in addition to the tangential velocity.

As in a previous paper (Abad et al. 2003, from now on [A03]), we present a
treatment of the proper motions based on Herschel's Method (Trumpler
\& Weaver 1953) and the Polinomio Deslizante (Stock \& Abad 1988).
Herschel's method reveals the existence of an apex when a
common motion is present. A simple equation relates proper motions
and the space velocity of the group, through the angular distance
from the apex to the stars' position on the celestial sphere, when
parallaxes are known. The Polinomio Deslizante (from now on, PD) is a
numerical routine that is able to detect systematic trends in a
sample of data, which, in our particular case are proper motions. 
Most previous work on the local kinematics using proper motions (and radial
velocities in some cases) usually employed a statistical analysis and
adopted an analytical model for the velocity distribution to which
proper motions are fitted, through a given set of parameters.
The Polinomio Deslizante has no assumptions about the data or their
distribution, so its results represent more faithfully the
observed reality, not being forced to follow a given equation.
A separate mention must be made of Dehnen (1998), who uses a non-parametric
maximum penalized likelihood method to estimate the velocity
distribution $f(v)$ of nearby stars, based on the positions
and tangential velocities of a kinematically unbiased sample of 14,369 stars 
selected from Hipparcos catalogue. He found that $f(v)$ shows a rich
structure in the radial and azimuthal motions, but not in the
vertical velocity. Some of these structures are related to well-known
moving groups.

Herschel's original idea was to use the Convergent Point Method to
determine the solar motion. Later it was used as an additional tool
to confirm the membership of stars in clusters and associations. 
In this paper we show that this method has more useful features.
The substitution of great circles, which represent the stellar
proper motions, by their corresponding poles, allows the matching
of both position and proper motion with a point on the celestial
sphere, for each star. In the case of a stellar cluster, the poles
of the cluster members are located on arcs of a great circle over
the celestial sphere, but if the apex is located inside the cluster or
the stars members are spatially distributed around the Sun, then the
poles span along a whole great circle.
Using the poles it is possible to extend Herschel's method from small
fields to large and dense catalogues, representing the proper motions 
in a more interesting way than the classical Vector-Point Diagram,
because the field of re\-pre\-sen\-ta\-tion is now the whole celestial
sphere. The denser the stellar field is, the more crowded the poles are,
but this representation highlights in a visual way, important
characteristics of the common motion, if there is any in the sample
studied.

Stellar associations, clusters, streams and any kinematically bound
group in general, can give us some information about the local (and
even the global) dynamical structure of the Galaxy. The main interest of this
paper is to demonstrate a procedure to detect and compute the motion of any
of these possible types of common stellar motions, by looking for
systematic trends in the proper motions of many stars,
covering large areas on the sky. Using this method on the Hipparcos
catalogue, we have found two large-scale patterns of motion, both
of them constrained to the galactic plane. One motion is directed
radially inwards to the galactic center, while the other is directed radially
outwards. For each of them, the apex and velocity are measured.

As Hipparcos stars are near the Sun, it can be difficult to associate
these and others similar ``streams'' to a specific origin. 
A list of well studied dynamical perturbations that could affect the solar
neighborhood, possibly linked to these patterns of motions, includes
for example, the Outer Lindblad Resonance (OLR) located just inside
the solar radius and produced by the galactic bar, according to
Mulbahuer \& Dehnen (2003). Another example, is the ongoing
disintegration of the Sagitarius dwarf galaxy, which is crossing the
galactic plane as it is being accreted by the Milky Way, at a distance
of only one kpc from the Sun (Majewski et al. 2003). The proximity of
the Sagitarius galactic arm can be listed as well. Their effects on the
local galactic kinematics are not quantitatively well known, although
some qualitative insights can be determined.

\section{The Polar Representacion of the Convergent Point Method (CPM)}

The great circle defined for each star, by its position and proper motion vector, 
has been used as a representation of the proper
motion in several investigations.  For example Schwan (1991),
selected stars of the Hyades open cluster and computed its apex based
on the great circles and their intersections. Both, Agekyan \& Popovich
(1993) and Jaschek \& Valbousquet (1992), used the CPM to determine the
solar motion from different catalogues and numbers of stars.
More recently Chereul et al. (1998) determined the space velocity
distribution of A-F stars in the Hipparcos data. 

In Abad (1996), the intersections between great circles was used
as a tracer of the existence of stellar clusters. The study of the 
density distribution of the intersections helped him to get the
apex of the Coma Berenices open cluster. A more mathematical treatment
on a global proper motion convergence map has been made by Makarov et al.
(2000), working on a sample of X-ray stars from the ROSAT All-Sky
Survey Bright catalogue. We show in this paper that the CPM, through
the polar representation of the great circle (see [A03]),
has more benefits to offer. 

Each great circle is uniquely defined by its pole $\vec{p}$, computed
by the cross product of the initial $\vec{x_i}$ and final position $\vec{x_f}$ of the star,
using the proper motion according to the following equations:
\begin{eqnarray}
\vec{x_i} &=& (\cos{l}\cos{b},\sin{l}\cos{b},\sin{b}) \\
\vec{x_f} &=& (\cos{(l+\mu_l\Delta t)}\cos{(b+\mu_b\Delta t)}, \nonumber \\ 
          & & \;\sin{(l+\mu_l\Delta t)}\cos{(b+\mu_b\Delta t)},\sin{(b+\mu_b\Delta t)}) \\
\vec{p} &=& \vec{x_i} \times \vec{x_f}
\end{eqnarray}
with $\Delta t = 1$ year. Figure 1 gives a comparative view of both, the CPM and the
Vector-Point Diagram, when applied to the Hipparcos catalogue.
To illustrate the CPM, we use the polar representation. 
This figure shows the existence of an area on the celestial
sphere with a major concentration of poles. Given that it is 
possible to fit a great circle to the data, the interpretation of this
plot may be related to the existence of a systematic component of
the stellar proper motions in the Hipparcos catalogue.
In all plots of the whole celestial sphere in this paper, galactic
longitude runs from $l=0^o$ to $l=360^o$ in order to make it easier
to visualize the results.

\begin{figure*}
\begin{center}
\includegraphics[width=15cm,height=8cm,angle=0]{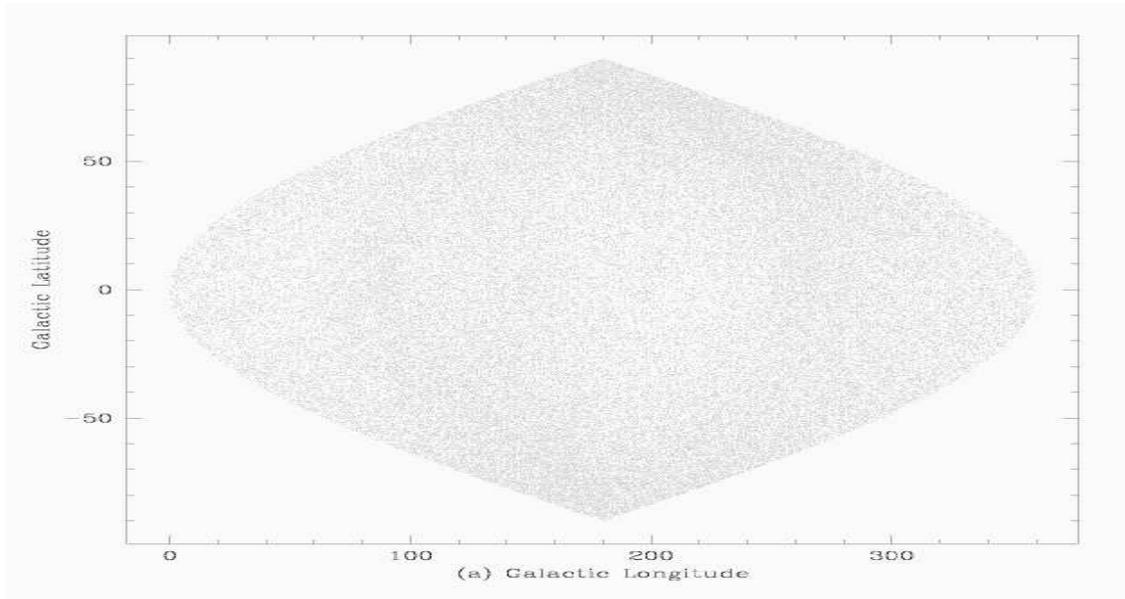}
\caption{Polar representation of the CPM for the same
stars as Fig. 1, once proper motions have been corrected 
for solar motion and differential galactic rotation. 
Therefore, these proper motions 
are referred to the Local Standard of Rest (LSR). 
When compared to Fig. 1, it is still possible to see the
dense area of poles but with a small change of orientation.}
\end{center}
\end{figure*}

\section{Solar Motion and Galactic Rotation}

Two well known sources of systematic motion, 
solar motion and differential galactic rotation,
are shared by all the stars around the Sun.

The shape of the feature observed in Fig. 1, made us think that
it was a reflex of the solar motion. Actually, this idea was
already studied in [A03], using the CPM+PD as we propose here.
In their paper, solar motion is obtained as
the common trend of motion shared by all stars in Hipparcos
Catalogue. As their method requires parallaxes,
only stars with $3\sigma_{\pi}\leq\pi$ were selected,
to avoid further problems. Their results are 
$(l_\odot,b_\odot)=(61^o37,19^o95)\pm (3^o13,2^o62)$
and $V_\odot=21.96\pm 3.72$ km/s. It is well known
from all papers about solar motion,
that the results depend on the stellar sample used.
For example, considering stars of only a given spectral type,
and comparing results between different samples,
reveals what is known as the ``asymmetric drift''.
As the true solar motion
is not responsible for this, then these results are basically
biased by how the samples were defined.

Any method to find the solar motion, rests on
the ability to find a stellar sample with no 
significant kinematic deviation with respect to the to the 
Local Standard of Rest (LSR).  Particularly in [A03], 
the solar motion was obtained based on what the PD detected 
in the whole sample of Hipparcos stars, regardless of spectral 
type and only limited by having good parallaxes.
Precisely the fact that solar motion is present in all
stellar proper motions, regardless of any condition,
justifies such a choice.

After correcting for both, galactic rotation and solar motion,
we are then in a reference frame where all stars have zero common motion.
There is no guarantee at all that this ``zero common motion 
reference frame'' has no net motion with respect to the LSR, 
but this is in our opinion the closest realization we have of the LSR,
based exclusively in data and with no theoretical assumptions.

Using [A03] values of $(l_\odot,b_\odot)$ and $V_\odot$, the effect
of the solar motion in the stellar proper motion, $(\mu_{l,\odot},\mu_{b,\odot})$
for a star with position $(l,b)$ is given by:
\begin{eqnarray}
\mu_{l,\odot}\cos{b} &=& \frac{\pi}{4.74}(u_\odot\sin{l}-v_\odot\cos{l})\label{s1}\\ 
       \mu_{b,\odot} &=& \frac{\pi}{4.74}(u_\odot\cos{l}\sin{b}+v_\odot\sin{l}\cos{b}\nonumber\\
                     & & -w_\odot\cos{b})\label{s2}\\
(u_\odot,v_\odot,w_\odot) &=& V_\odot\; (\cos{l_\odot}\cos{b_\odot},
\sin{l_\odot}\cos{b_\odot},\sin{b_\odot}) 
\end{eqnarray}
where $(u_\odot,v_\odot,w_\odot)$ are the
rectangular components of the solar motion in km/s, directed towards
the galactic center, the direction of galactic rotation and the galactic north pole,
respectively. $\pi$ is the parallax of the star. 

Differential galactic rotation also produces part
of the observed proper motions, introducing a systematic 
component which depends mostly on the position in the galaxy,
but it is independent of the distance to the star. Oort's constants
$A$ and $B$ are used to make simple models of the differential galactic
rotation, although more sophisticated methods (Mignard 2000) 
express the galactic rotation in terms of the Generalized Oort's constants. 
We particularly consider equations (27) and (28) of Mignard (2000),
to model the differential galactic rotation:
\begin{eqnarray}
\mu_{l,gr}\cos{b} &=& \bar{A}\cos{b}\cos{(2l-2\phi)}+A'\sin{b}\cos{(l-\psi)} + \nonumber\\
		  & & B\cos{b}-B'\sin{b}\cos{(l-\chi)} \\
       \mu_{b,gr} &=& -\frac{\bar{A}}{2}\sin{2b}\sin{(2l-2\phi)}+A'\cos{2b}\sin{(l-\phi)}\nonumber \\
		  & & -\frac{K}{2}\sin{2b}+B'\sin{(l-\chi)} 
\end{eqnarray}
where the values of the Generalized Oort's constants
$\bar{A}, A', B, B'$ and $K$, which depend on the spectral type,
are listed in Table 5 of Mignard (2000).

Both $(\mu_{l,\odot},\mu_{b,\odot})$ and $(\mu_{l,gr},\mu_{b,gr})$
are subtracted from the individual stellar proper motions. Figure 2
shows the polar representation of the proper motions after they
are corrected for these two effects. This plot represents
the intrinsic motions of the stars, referred to the LSR.
Once again, it is still possible to see the dense
area of poles making a wide band surrounding the celestial sphere,
but with a small change of orientation when compared to Fig. 1. 
This tells us two new things: (a) solar motion was not the origin,
and (b) it is not a common systematic trend for the totality
of the stars in the catalogue, but for a subsample. It is
necessary to stress that the differential galactic rotation model
of Mignard (2000), does not produce the great circle of poles like
the one found in Fig. 2.

\section{Selecting stars with common space motion}
Figure 2 shows clearly a band with an enhanced density of poles.
The possibility of fitting a great circle would imply the existence
of a common space motion. The width of the band 
represents the internal dispersion around the mean motion of the
group. Of course, proper motion errors also play a role and
another factor could be a possible selection effect of the Hipparcos
mission, although we later prove that this band corresponds to
a real space motion. 

Not all of the stars are physical members of this wide great circle,
therefore the first step is to select the stars on it and
search for possible members to get the modulus and
direction of their motion. A half-width of $14^o$ was chosen
around the best fitting central great circle. If we think
of this great circle as an ``equator'', then the corresponding
``poles'' ($(l_0,b_0)$ and its opposite) are probable apexes
of the motion. Then, it is necessary to separate the stars producing
the band into two samples, depending on the direction of their
proper motions: those pointing towards $(l_0,b_0)$, hereafter
Sample 1, and those pointing the other way, hereafter
Sample 2. Then, making use of parallaxes, the corrected proper
motions are re-scaled to a fixed distance of 100 pc and we
apply the PD fitting function on them, in a si\-mi\-lar way
to that done in Abad et al. (2003). If there is a common space
motion in these stars, the re-scaled proper motions $\mu_{scaled}$ 
have a common behavior, that can be detected and measured by
the PD.

For any given point on the sky, the PD produces a vector
which represents the common component, shared by all the
stars surrounding the point up to a given predefined
radius. Therefore two points separated by twice this radius
have output vectors completely independent, given that the
proper motion data contributing to their computation are
distinct. To easily visualize global and/or local patterns
in the PD resulting vectors, we define an evenly distributed
grid of points all over the sky. Working on Sample 1 for
example, after a first application of the PD, a global
vectorial pattern of proper motions is found. To verify
if it corresponds to a common space motion, all following
conditions have to be fulfilled:
(a) the pattern must be smooth and systematic all over the sky, 
(b) both apex and antapex must exist, and 
(c) the modulus of the vectors obtained with
the PD must roughly follow Hershel's formulation, meaning
that the modulus of the vectors depend on the angular distance
to the apex, being maximum at 90 degrees from it and close to
zero in both apex and antapex.

To proceed later with Herschel's method, we define another
grid of points, this time making meridians and parallels 
as if $(l_0,b_0)$ and its opposite were the poles of the
sphere. The PD is applied on these points and conditions
(a), (b) and (c) are checked again. Results of this procedure
are shown in Fig. 3 for Sample 1. A clear global pattern is
observed. Zooming in on the area of minimum-modulus vectors
(Fig. 4c) helps us to select an apex. Additionally, the CPM
poles of the PD vectors, fall in a quite narrow great circle
that is fitted and from this we measure a better apex.

For each meridian of the grid defined by the apex, we plot
the PD vectors modulus $\mu_{PD}$ versus their position along
the meridian (equivalent to the angular distance $\lambda$ from
the vector position to the apex). Actually, we transform those
moduli to their corresponding tangential velocity in km/s, using
$V_t=4.74\mu_{PD}\; 100.0$ given that all proper motions (in ''/yr)
were rescaled to a distance of 100 pc.
Figure 4a plots the corresponding value of $V_t$ for all vectors
of Fig. 3, distributed by meridians. According to Herschel's
formulation, for a perfect common space motion, each
meridian satisfies $V_t=V\sin(\lambda)$, 
where $V$ is the total space velocity of the motion,
and those vectors located at $\lambda=90^o$ have $V_t=V$.
Figure 4a, can certainly be fitted by a sine function.

Simultaneously, the modulus of the vectors should
remain constant for each parallel of the grid.
The constant modulus of the vectors located on the equator of the grid,
should have the maximum value, because all of them
are $90^o$ from the apex, and their corresponding $V_t$
matches the space velocity $V$ of the motion. Figure 4b shows $V_t$
along the different parallels in the grid, for the
same proper motion vectors as in Fig. 3.

Finally, based on the $V$ value obtained from the sine function fitted
in Fig. 4a, we make a histogram of the 
ratio $R=(4.74\mu_{scaled}\; 100.0)/(V\sin(\lambda))$
for each star, as shown in Fig. 4d.
The true $V$ of the motion should produce a 
peak in the histogram around $R=1$,
though a slightly different value will just shift
the position of the peak. More important is the fact that
clear outliers can be easily rejected, for example
in Fig. 4d those stars with $R>4$ were eliminated.
In addition to this ``comparison of length'' 
we also check the angle between the observed $\mu$
and the {\it kinematics proper motion} $\mu_{kin}$ the star should have,
if it were perfectly directed towards the apex with an exact
velocity of $V$. Those stars with very large angles 
were rejected. The purpose of this cleaning is just
to get rid of outliers.
A new apex is then calculated and the process is iterated.

\begin{figure*}
\begin{center}
\includegraphics[width=15cm,height=8cm,angle=0]{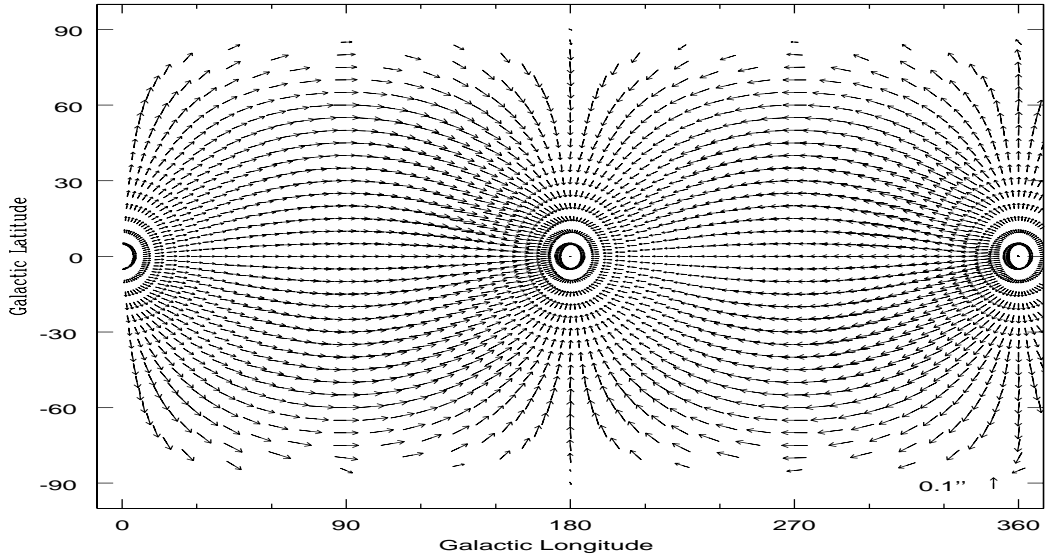}
\caption{Vectors of proper motions produced by the Polinomio
Deslizante, applied to those stars whose paths are within a $14^o$
radius centered on $(l,b)=(180,0)$ and taking this point as the apex.
This was the first pre-selection of stars done for Sample 1.}
\end{center}
\end{figure*}

\begin{figure*}
\begin{center}
\includegraphics[width=15cm,height=8cm,angle=0]{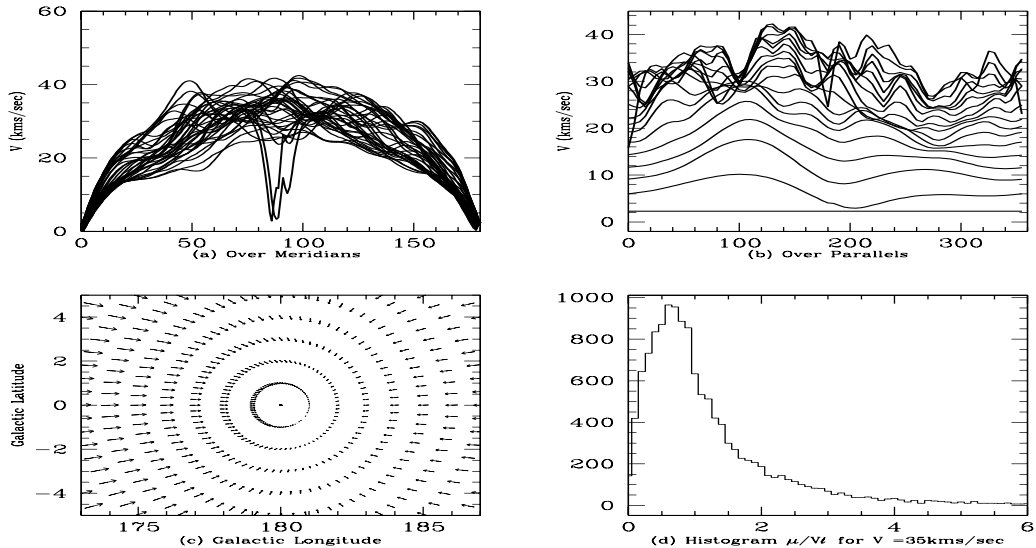}
\caption{a) $V_t$ along meridians are similar and can be fit by
a sine function of the angular distance $\lambda$ to the apex. b)
$V_t$ along parallels. c) An apex point with a positive increment
on longitude and negative increment on latitude is closer to the
true apex. d) The histogram indicates that the real total space
velocity is lower than the initially chosen. Irregular values
around $\lambda=90^o$ come from trigonometric singularities for
proper motion vectors located close to the galactic poles
$b=\pm 90^o$.}
\end{center}
\end{figure*}

\begin{figure*}
\begin{center}
\includegraphics[width=15cm,height=8cm,angle=0]{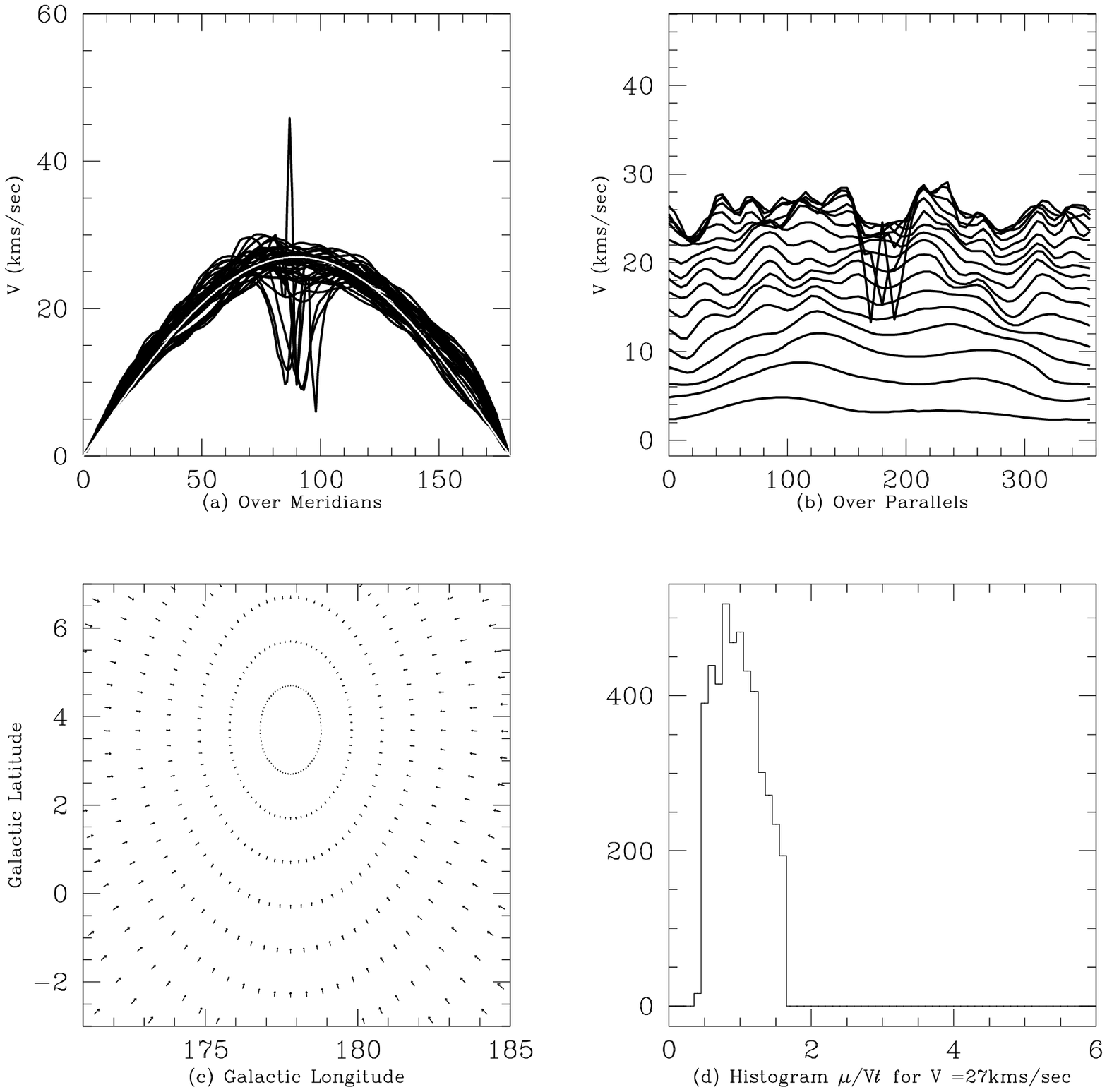}
\caption{Final results for Sample 1. The motion is directed
to $(l,b)=(177^o8,3^o7)$ with a space velocity of V=27.2 km/s.
Irregular values around $\lambda=90^o$ come from
trigonometric singularities for PD vectors located 
close to the galactic poles $b=\pm 90^o$.}
\end{center}
\end{figure*}

\begin{figure*}
\begin{center}
\includegraphics[width=15cm,height=8cm,angle=0]{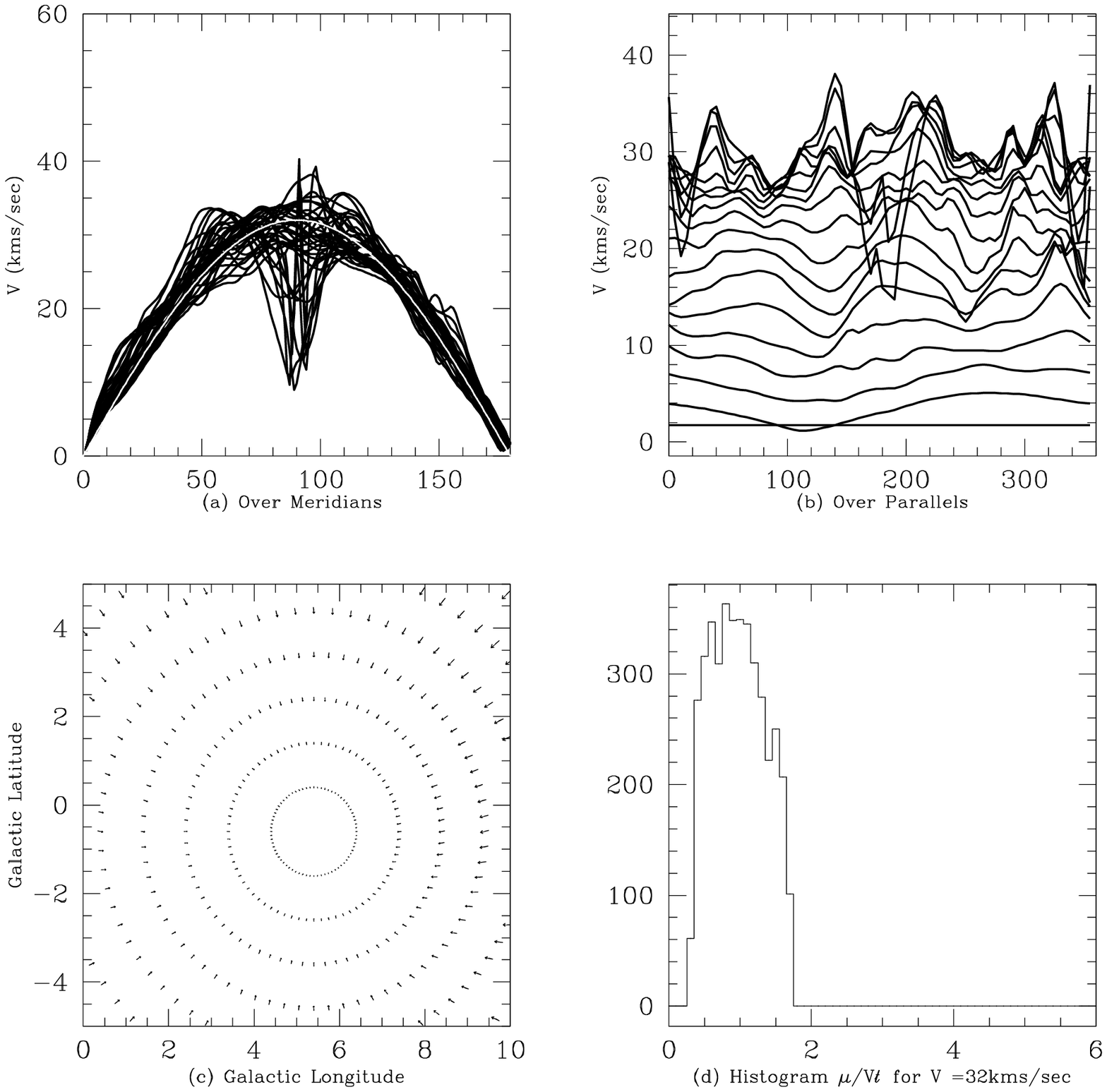}
\caption{Final results for Sample 2. The motion is directed to 
$(l,b)=(5^o4,-0^o6)$ with a space velocity of V=32.1 km/s.
Irregular values around $\lambda=90^o$ come from
trigonometric singularities for PD vectors located 
close to the galactic poles $b=\pm 90^o$.}
\end{center}
\end{figure*}

\section{Two preferential patterns of motion}

The first preselection of stars, done as explained at the
beginning of Sect. 4, consisted of 24486 stars from Hipparcos 
catalogue, 13208 with proper motion towards (l,b)=(180,0),
named Sample 1, and 11278  with proper motion towards 
(l,b)=(0,0), named Sample 2. Stars with $3\sigma_{\pi}>\pi$ 
and Hyades cluster members had
previously been rejected. We will see later that our final
results do not change, even including the Hyades cluster (which has
a strong and well defined motion).

\subsection{Sample 1}
Figure 5 shows the final results for Sample 1, after a few iterations
of the procedure explained in Sect. 4.
The number of selected stars is now reduced to 4566, and they produce
a coherent systematic pattern of motion with apex $(l,b)=(177^o8,3^o7)\pm
(1^o5,1^o0)$ and velocity $V=27.2\pm 1.1$ km/s. Sample 1 stars are
located all around the sky and do not exhibit any particular
preferred condition on position, distance (space distribution
(X,Y,Z), Fig. 8), color, magnitude or spectral type 
(color-magnitude diagram, Fig. 7). These stars do not display any
noticeable difference with the general distribution of Hipparcos stars.

\subsection{Sample 2}
Figure 6 shows the final results for Sample 2. The number of selected
stars is now reduced to 4083, and they produce a coherent systematic
pattern of motion with apex $(l,b)=(5^o4,-0^o6)\pm (1^o9,1^o1)$ and
velocity $V=32.1\pm 2.1$ km/s. Sample 2 stars, like those of Sample 1,
do not exhibit any particular preferred condition (Fig. 7), except
for a $45^0$ tilted preferential line on the (X,Y) plane and a slightly tilted
orientation in the (Y,Z) plane (Fig. 8).

\begin{figure}[h!]
\begin{center} 
\includegraphics[width=8cm,height=8cm,angle=0]{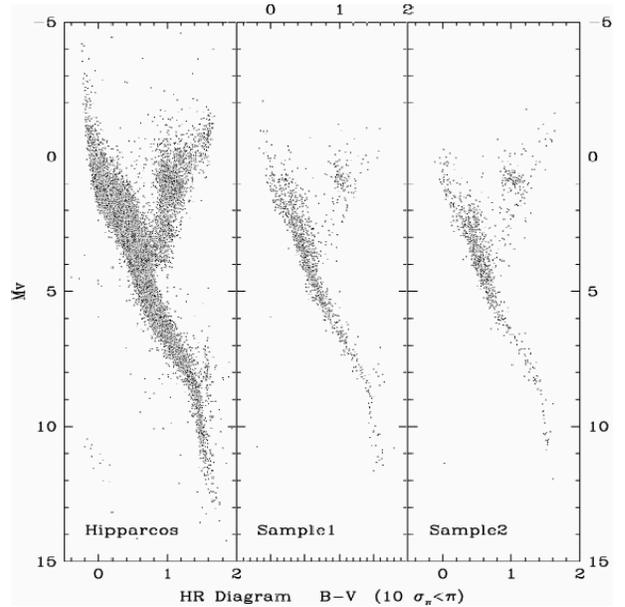}
\caption{Color-Magnitude Diagram for Hipparcos catalogue and Samples
1 and 2, including only stars with $3\sigma_{\pi}<\pi$. Distributions
are similar.}
\end{center}
\end{figure}

\begin{figure}[h!]
\begin{center} 
\includegraphics[width=8cm,height=8cm,angle=0]{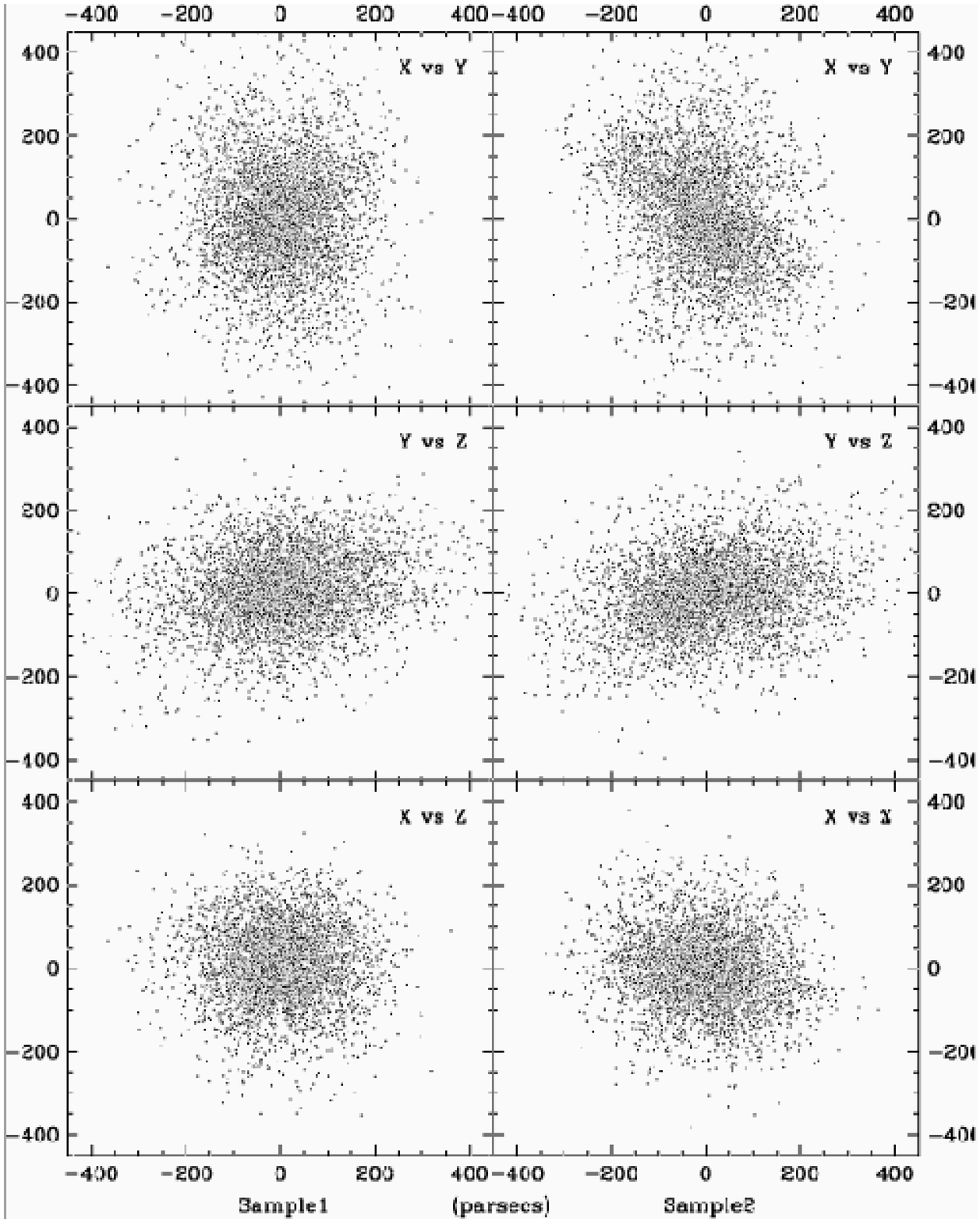}
\caption{Spatial distribution (X,Y,Z) for Sample 1 and 2. There is a
$45^o$ tilted (X,Y) preferential line and a (Y,Z) slightly tilted
orientation for stars of Sample 2.}
\end{center}
\end{figure}

\section{Cross-checking the results}

The space motion of a star is the vector sum of its tangential
and radial components. If available, radial velocities
can be used as an independent test of the results
obtained using only proper motions. This is true, given
that for a common space motion
$V_r=V\cos(\lambda)$ so that radial velocities complement 
tangential velocities for each star.

With this purpose in mind, we use the ``Catalogue G\'en\'eral de
Vitesses Radiales Moyennes pour les Etoiles Galactiques'' 
(Barbier-Brossat \& Figon 2000). The catalogue contains
mean radial velocities for 36145 stars of which 1198 were
matched with final Sample 1 and 886 with final Sample 2.
Figure 9 plots for each sample, radial velocity versus
galactic longitude, once corrected for solar motion.

The existence of a tendency in the radial velocities, which properly
complements the patterns found in the proper motions, is
evident for Sample 1. Extreme and opposite values around $\pm 27$
km/s, are found close to $l=0^o$ and $l=180^o$. Minimum values of 0
km/s are observed around $l=90^o$ and $l=270^o$. A similar, but
noisier trend is observed for Sample 2.

\begin{figure}[h!]
\begin{center} 
\includegraphics[width=8cm,height=8cm,angle=0]{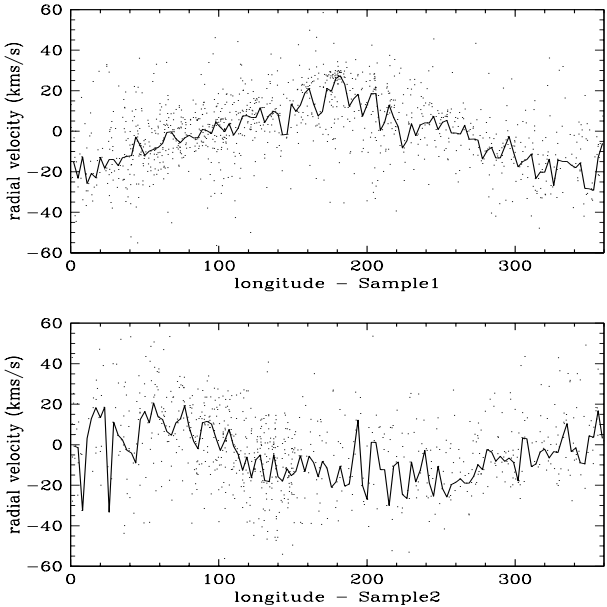}
\caption{Distribution of radial velocities vs. galactic longitude
for those stars of the ``Catalogue G\'en\'eral de Vitesses Radiales
Moyennes pour les etoiles galactiques'' included into
Samples 1 and 2. 1198 stars belong to Sample 1 (upper figure) and
886 stars to Sample 2 (lower figure). In both cases
the radial velocities have been corrected for solar motion. Median
values over $4^o$ longitude interval are plotted with a solid line.
Excellent agreement with the results shown in Fig. 5 exists for Sample 1.}
\end{center}
\end{figure}

\section{Cross-checking the method}

With the only aim of checking the correctness and robustness of the method,
we apply the PD+CPM poles to several ``dynamical streams'' detected
by Famaey et al. (2005) (from now on [F05]). 
In their paper, a maximum-likelihood method
based on a Bayesian approach is used to derive the kinematic properties
of several subgroups. A sample of around 6000 giant stars, with
proper motion, parallax and radial velocity data (Tycho-2, Hipparcos and CORAVEL),
are used to reconstruct 3D space velocities of the stars.

Figure 11 shows the polar representation of the different 
stellar streams and groups as listed by [F05].
the Young Giants, the Hyades-Pleiades supercluster, 
the Hercules stream, the Sirius moving group and the
so-called Smooth Background. The proper motions
from which these poles were computed, are already corrected
for differential galactic rotation and solar motion,
using the values and equations cited in [F05], as explained
in next parapraph. 
These corrections assure us that whatever the PD detects later
as a common motion in any of these samples, 
corresponds to a real intrinsic spatial motion and not to
a common component induced by some external effect.
We first use [F05] values to be consistent with their
work, although we later try [A03] values as well.

Following [F05], given the observed proper motions
$\mu_{l,0}$ and $\mu_{b,0}$, the corrected values for
galactic rotation are:
\begin{eqnarray}
\kappa\mu_l &=& \kappa\mu_{l,0}-A\cos{2l}-B \\
\kappa\mu_b &=& \kappa\mu_{b,0}+1/2 A\sin{2b}\sin{2l} 
\end{eqnarray}
where $\kappa=4.74$ is the factor to convert proper motions
into space velocities. $A=14.82$ km/s/kpc and 
$B=-12.37$ km/s/kpc are the Oort constants,
derived by Feast \& Whitelock (1997). The effect of solar motion
on the proper motion of the stars is computed following
equations (\ref{s1}) and (\ref{s2}), with
$(u_\odot,w_\odot)=(10.24,7.77)$ km/s, as measured by [F05].
Knowing that velocities along the direction
of galactic rotation are affected by the asymmetric drift, 
[F05] adopted the value of $v_\odot=5.25$ km/s
from Dehnen \& Binney (1998).

It is straightforward to see that the stars in the
Young Giants group, do not exhibit a particular
trend in their poles, except some accumulation 
near the galactic poles. This is consistent with these stars
being young disk population. Similarly, the High-Velocity
stars (not shown in Fig. 11), have their poles more or
less randomly distributed across the celestial sphere.

\begin{figure}[h!]
\begin{center}
\includegraphics[width=8cm,height=8cm,angle=0]{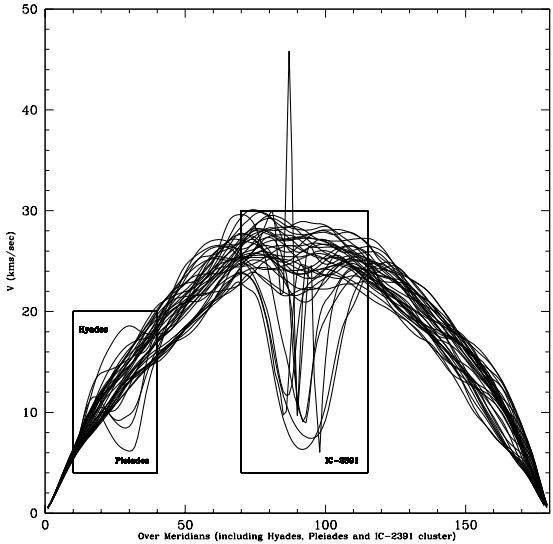}
\caption{Local deformations to Fig. 5a appear when the Hyades,
IC-2391 and Pleiades clusters are included into Sample 1.}
\end{center}
\end{figure}

\begin{figure*}
\begin{center}
\includegraphics[width=15cm,height=9.25cm,angle=0]{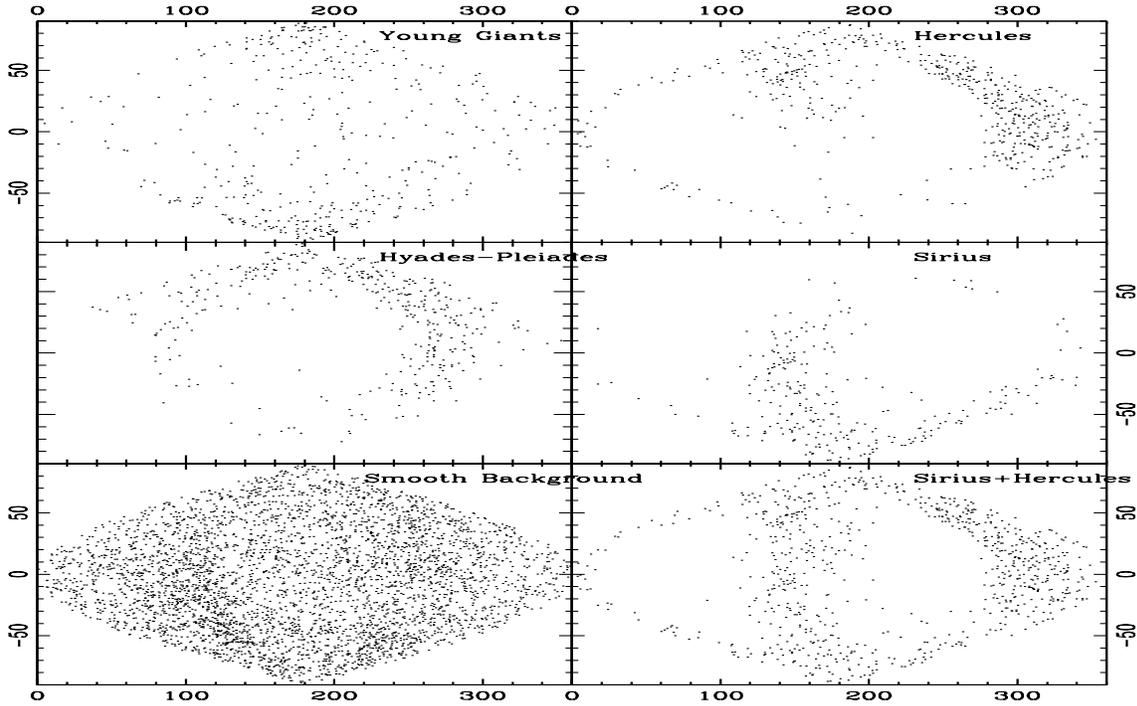}
\caption{Polar representation of the ``dynamical streams'' of Famaey et
al (2005), in galactic coordinates, already corrected for differential
galactic rotation and solar motion.}
\end{center}
\end{figure*}

On the other hand, the Hyades-Pleiades supercluster's poles are
located on a band, close to the wide great circle of poles
detected by us (see Fig. 2). Even more, our pattern is also
visible in the poles of the stars of the Smooth Background,
which according to [F05], corresponds to an
``axisymmetric'' mixed population of the galactic disk,
composed of stars born at many different epochs.

The Sirius and Hercules streams deserve special attention, since each 
have their poles located along and around the arc of a great circle,
clearly biased because [F05] only use stars from the
northern hemisphere. The lower right panel of Fig. 11 shows that both
distributions may be complementary. Based on these distributions it is
clear that both motions are almost opposed and close to the galactic plane.
We most stress here that we are not relating Sirius and Hercules streams
to the pattern of motions we found in Hipparcos.
What is really interesting for us is that the distribution 
of their poles, as seen in Fig. 11 right lower panel, 
also lies on a great circle. 
This offers us the opportunity to test our method and verify that it
is able to detect, separate and measure their corresponding apex and velocity.

After correcting for galactic rotation and solar motion,
the proper motions are re-scaled to a distance of 100 pc
and the PD is applied. For this, a grid of points separated by
10 degrees is defined and we set a $15^o$ radius around each point
to guarantee a solution from PD routine. The PD vectors found are
shown in the upper panels of Fig. 12. They follow the three conditions
stated in Sect. 3, indicating the possible existence of a 
intrinsic spatial common motion.

\begin{figure*}
\begin{center}
\includegraphics[width=15cm,height=9.25cm,angle=0]{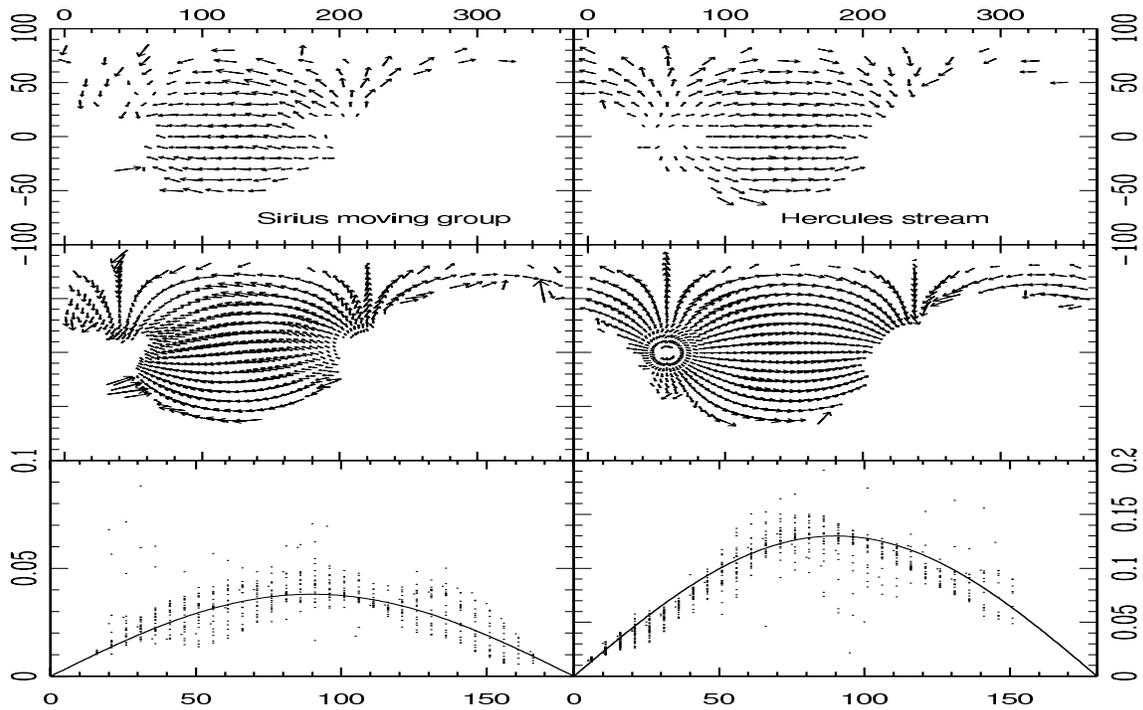}
\caption{Upper and middle panels: vectorial representation of the PD
results, in galactic coordinates, for the Sirius and Hercules streams.
Lower panels: PD vectors' moduli ("/yr) versus apex-position angle.
A sine function is plotted with a solid line.}
\end{center}
\end{figure*}

For an initial determination of modulus and apex of both motions,
a new grid of points evenly distributed along meridians
and parallels is defined, so that apex and antapex are the new
poles of the sphere. Once the PD is applied to the 
new points (Fig. 12 middle panels), we check that the vectors'
moduli follow Herschel's formulation (lower panels Fig. 12).
Initial results indicate the existence of a common motion,
to which we assign a value for the Hercules
stream of: apex $(l,b)=(237.5,-0^o5)$, $V=61.5$ km/s and for 
Sirius stream: apex $(l,b)=(40,-6^o)$, $V=18.01$ km/s, when taking,
respectively, the values of $\mu=0.13$ "/yr and $\mu=0.038$ "/yr as the 
scaling factors for the sine functions.
According to [F05], the Hercules stream has an apex $(l,b)=(230^o8,-0^o1)$
with $V=67.1$ km/s, and the Sirius stream has an apex $(l,b)=(58^o8,-34^o1)$
with $V=9.6$ km/s.

It is necessary to remark that both samples are contaminated,
in our opinion, by stars that do not follow the stream's general trend.
Nevertheless, it is not our aim to study them but rather to show that 
our method is able to detect the streams within a sample that certainly 
contains them. Our results are clear and stable, and 
they differ numerically from those cited by [F05], just because:
(a) We correct the stellar proper motions for solar motion 
and [F05] does not, and (b) different methods are employed to measure 
the apex and the velocity.  

We additionally made this same cross-checking, using the [A03] solar motion 
and the Mignard (2000) model of galactic rotation, and we again recover
both streams.  Differences between 5 and 10 km/s in $V$ appear,
due to the quite different model of solar motion applied. Some
smooth corrections also appear for the apex, but they continue
near the galactic plane.

Returning to our data, the strength of the method is 
certainly proved, when the inclusion into the first preselected
Sample 1, of data from known clusters with well defined motions,
produces only smooth local but no general variations to the final
results. Figure 10 shows the same kind of plot as Fig. 5a,
when the Hyades, IC-2391 and Pleiades clusters are included. Similar
deformations are introduced in Fig. 6a, when the Sirius supercluster
is included as well.

\begin{figure*}
\begin{center}
\includegraphics[width=15cm,height=8cm,angle=0]{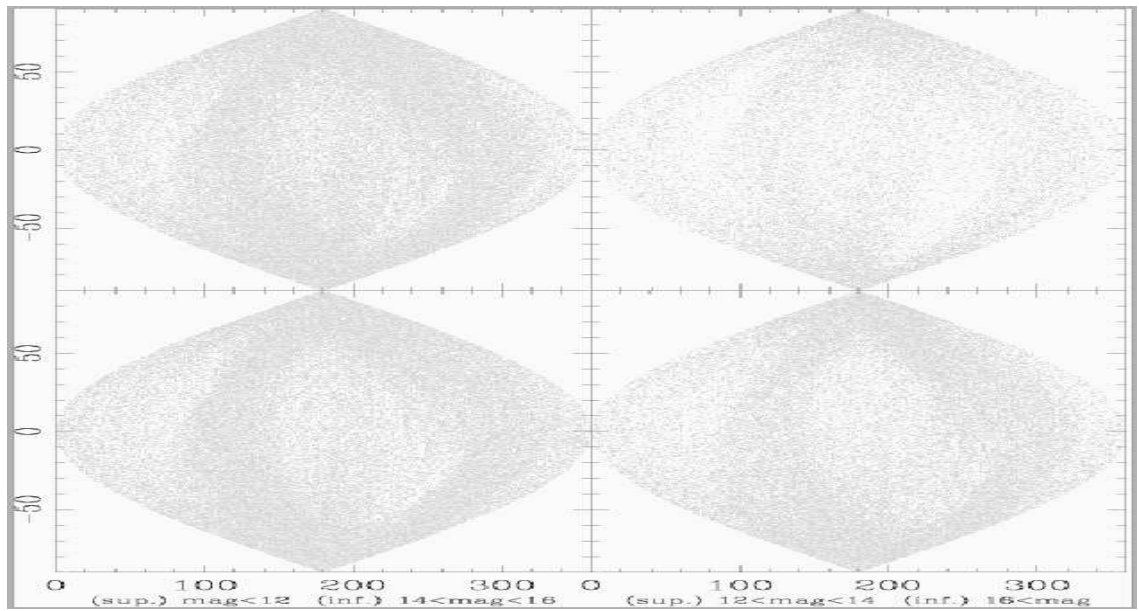}
\caption{Polar representation of the CPM for the NPM2 catalogue
in galactic coordinates, separated by range of magnitude. It is
possible to see a pattern similar to that found in Fig. 1 for the
Hipparcos catalogue, which motivated this paper.}
\end{center}
\end{figure*}

\section{Discussion and Conclusions}

In this paper we show a procedure that is able to detect
trends of motion in samples of stars. Using proper motions
and parallaxes, but not requiring 
radial velocities, we can obtain the modulus and direction
of the motion.  Therefore, it has the capability 
to detect stellar associations, moving groups and
streams, particularly when working on extended areas.

Stellar streams have been known from many years.
Between 1925 and 1930, Ralph Wilson and Harry Raymond
with proper motions, and Gustav Str\"{o}mberg with radial 
velocities, dealt with the two Kapteyn streams and
other features in the velocity distribution.
Streams were a main subject of study by O.J. Eggen (1996a, 1996b), 
who located some of them using (U,V) velocities
preferentially located in the second quadrant.
Gozha (2001) made a study of Eggen's Groups, indicating
that Stream I (associated with the Hyades cluster), and Stream II
(Pleiades and Sirius Supercluster) are the most crowded ones.

The knowledge of distance, position, motion and the probability of
membership for known stellar clusters, 
helps study the streams associated with them, but on the other hand, 
it makes the clusters an indispensable
tool to begin with. The initial criterion we use to select stars 
as probable members of a common motion depends exclusively on
the individual proper motion. We choose stars whose paths cross
a preselected area, with no assumptions about space location or gravitational conditions
(such as being linked to a known cluster). In this way it
is possible to detect patterns of motion that make a portion, but not
the total of the stellar motion. Certainly, this area is not chosen 
randomly, polar representation of
CPM and the PD numerical routine help us in this selection.

As our method makes use of the parallaxes, this means
that Hipparcos is the most extensive catalogue so far, to which it
can be applied. Nevertheless, some additional evidence,
supporting the fact that these patterns are extended,
can be obtained from other databases, like the NPM2 catalogue
(Hanson et al. 2003). This catalogue contains absolute proper
motions for 232000 stars north of $\delta=-23^o$,
in a blue apparent magnitude range from 8 to 18. The ca\-ta\-lo\-gue
covers $28\%$ of the northern sky, lying near the plane of
the Milky Way and many of the stars are included in fields away
from the galactic plane. Figure 13 shows the equivalent to Fig. 1, using
NPM2 original proper motions. The clear presence of an enhanced
density zone of poles along a wide great circle, motivated this paper.

Detecting and measuring the different kinematic populations
in the solar neighborhood is of the foremost importance.
The structure of the local velocity distribution reveals
information about the dynamic equilibrium of the solar
neighborhood and the galaxy.  Assuming the predominance of
the global over the local galactic potential, the disturbing 
effect of a non-axisymmetric component, like the rotating bar,
could explain some of the ``large-scale'' motions that many
stars exhibit around us.  On the other hand, local and temporal 
disturbances of the potential could also generate the observed 
groups of common velocity.

A precise cause for the patterns of motion detected in this paper is
difficult to establish. For a discussion of the possible origins
for such streams, refer to [F05].
Further observations are required to 
accurately sample the solar neighborhood velocity field.
On the other hand, pushing the distance limits outwards in the
astrometric catalogues, makes it imperative to create suitable procedures
in order to properly manage large amounts of data. Sampling a greatly
extended volume of space, surely means that several other forces and
motions come to play, and makes more difficult the detection of kinematically
bound groups of stars.

The goal of this paper has been to develop a
new way to study the kinematics of
the solar neighborhood, in the search
of streams, associations, clusters and any other 
space motion shared by a large number of stars,
without being restricted
by the availability of radial velocities.

\begin{acknowledgements}{}
We would like here to thank Dr. Jurgen Stock (1923-2004) 
for a full life of work and commitment to astrometry. This paper is dedicated
to him, to honor all his achievements, 
the work he started will continue well into the future.
We also deeply thank Bill van Altena (Yale University)
for help with the English version of this paper.
\end{acknowledgements}


\end{document}